\documentclass[10pt,twocolumn,letterpaper]{article}
\usepackage{times,graphicx,amsmath,amssymb,booktabs,array,multirow}
\usepackage[dvipsnames]{xcolor}
\usepackage{algorithm,algorithmic,microtype,colortbl,xspace}
\usepackage[numbers,sort&compress]{natbib}
\usepackage[format=plain,labelformat=simple,labelsep=period,font=small,compatibility=false]{caption}
\usepackage[font=footnotesize,skip=3pt,subrefformat=parens]{subcaption}
\usepackage[hyphens]{url}
\definecolor{linkblue}{rgb}{0.21,0.49,0.74}
\usepackage[breaklinks,colorlinks,allcolors=linkblue]{hyperref}
\hypersetup{%
  pdftitle={Streaming Generative Video Compression with Embedding Alignment and In-Context Propagation},
  pdfauthor={Zhitao Wang, Hengyu Man, Han Liu, Wenrui Li, Debin Zhao, Xiaopeng Fan}}
\usepackage[capitalize]{cleveref}
\crefname{section}{Sec.}{Secs.}
\Crefname{section}{Section}{Sections}
\Crefname{table}{Table}{Tables}
\crefname{table}{Tab.}{Tabs.}
\usepackage[shortlabels,inline]{enumitem}
\setlist[itemize]{noitemsep,leftmargin=*,topsep=0em}
\setlist[enumerate]{noitemsep,leftmargin=*,topsep=0em}

\renewcommand{\paragraph}[1]{\vspace{.08em}\noindent\textbf{#1.}}

\makeatletter
\def\@maketitle{%
  \newpage\null\vskip 4pt
  \begin{center}
    {\Large\bfseries\@title\par}
    \vskip 10pt
    {\large\lineskip .5em\begin{tabular}[t]{c}\@author\end{tabular}\par}
    \vskip 8pt
  \end{center}}
\renewenvironment{abstract}{%
  \centerline{\large\bfseries Abstract}\vspace*{8pt}\noindent\itshape\ignorespaces
}{\par\vspace*{8pt}}
\def\preprintsection{\@startsection{section}{1}{\z@}{-8pt plus -2pt minus -2pt}{5pt}{\large\bfseries}}
\def\preprintssect#1{\preprintsection*{#1}}
\def\preprintsect#1{\preprintsection{\texorpdfstring{\hskip -1em.~}{}#1}}
\def\section{\@ifstar\preprintssect\preprintsect}
\def\preprintsubsection{\@startsection{subsection}{2}{\z@}{-6pt plus -2pt minus -2pt}{4pt}{\fontsize{11}{13}\selectfont\bfseries}}
\def\preprintssubsect#1{\preprintsubsection*{#1}}
\def\preprintsubsect#1{\preprintsubsection{\texorpdfstring{\hskip -1em.~}{}#1}}
\def\subsection{\@ifstar\preprintssubsect\preprintsubsect}
\def\preprintsubsubsection{\@startsection{subsubsection}{3}{\z@}{-6pt plus -2pt minus -2pt}{3pt}{\normalsize\bfseries}}
\def\preprintssubsubsect#1{\preprintsubsubsection*{#1}}
\def\preprintsubsubsect#1{\preprintsubsubsection{\texorpdfstring{\hskip -1em.~}{}#1}}
\def\subsubsection{\@ifstar\preprintssubsubsect\preprintsubsubsect}
\DeclareRobustCommand\onedot{\futurelet\@let@token\@onedot}
\def\@onedot{\ifx\@let@token.\else.\null\fi\xspace}

\newcommand{\relaxexperimentspacing}{%
  \setlength{\textfloatsep}{18pt plus 2pt minus 3pt}%
  \setlength{\dbltextfloatsep}{18pt plus 2pt minus 3pt}%
  \setlength{\floatsep}{12pt plus 2pt minus 2pt}%
  \setlength{\dblfloatsep}{12pt plus 2pt minus 2pt}%
  \setlength{\intextsep}{12pt plus 2pt minus 2pt}%
  \def\preprintsection{\@startsection{section}{1}{\z@}{-10pt plus -2pt minus -2pt}{7pt}{\large\bfseries}}%
  \def\preprintsubsection{\@startsection{subsection}{2}{\z@}{-8pt plus -2pt minus -2pt}{5pt}{\fontsize{11}{13}\selectfont\bfseries}}}
\newcommand{\restorefloatspacing}{%
  \setlength{\textfloatsep}{20pt plus 2pt minus 4pt}%
  \setlength{\dbltextfloatsep}{20pt plus 2pt minus 4pt}%
  \setlength{\floatsep}{12pt plus 2pt minus 2pt}%
  \setlength{\dblfloatsep}{12pt plus 2pt minus 2pt}%
  \setlength{\intextsep}{12pt plus 2pt minus 2pt}}

\makeatother

\title{Streaming Generative Video Compression\\
with Embedding Alignment and In-Context Propagation}
\author{%
Zhitao Wang\textsuperscript{1},
Hengyu Man\textsuperscript{1},
Han Liu\textsuperscript{1},
Wenrui Li\textsuperscript{1},
Debin Zhao\textsuperscript{1},
Xiaopeng Fan\textsuperscript{1,2,3}\\[6pt]
{\small \textsuperscript{1}Harbin Institute of Technology}\\
{\small \textsuperscript{2}Harbin Institute of Technology Suzhou Research Institute}\\
{\small \textsuperscript{3}Peng Cheng Laboratory}\\[4pt]
{\small zhitao.wang.hit@outlook.com, manhengyu@hotmail.com, 24b903011@stu.hit.edu.cn,}\\
{\small liwr618@163.com, \{dbzhao, fxp\}@hit.edu.cn}}
\date{}

\begin{document}
\maketitle
\begin{abstract}
Video compression at ultra-low bitrates presents substantial challenges in preserving perceptual quality. Although pretrained video diffusion transformers (DiTs) provide powerful generative priors, existing generative video compression (GVC) methods built on them typically rely on computationally expensive VAE decoders to reconstruct video frames. Moreover, jointly refining multiple frames to exploit the inherent spatiotemporal modeling capabilities of video DiTs introduces additional buffering delay. To address these challenges, we propose SGVC, a streaming generative video compression framework that adapts a pretrained latent video DiT into a causal pixel-space refiner. Specifically, SGVC projects decoded features from a contextual pixel codec directly into the DiT input embedding space. To bridge the representation gap between codec-derived embeddings and pretrained DiT embeddings, we introduce Embedding Alignment, which aligns the projected embeddings with input embeddings extracted from the original frames during training. The refined DiT hidden states are then mapped directly to the pixel domain through a lightweight reconstruction path, bypassing the pretrained VAE decoder. To further exploit the spatiotemporal priors of the video DiT while maintaining causal decoding, we introduce In-Context Propagation, which reuses layer-wise K/V states across frames to exploit historical generative context, improving temporal consistency without requiring clip-wise buffering. Extensive experiments demonstrate that SGVC achieves competitive rate--perception performance, with average BD-rate savings of 71.97\% over S$^2$VC in terms of DISTS.
{\urlstyle{same}Project Page: \url{https://chigland.github.io/SGVC/}}
\end{abstract}

\section{Introduction}

\begin{figure}[t]
\centering
\includegraphics[width=1.0\linewidth]{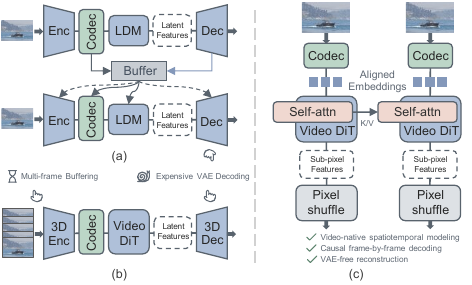}
\caption{
Comparison of diffusion-based video compression paradigms. (a) Framework based on image latent diffusion model (LDM) with VAE decoder. Frame-wise methods employ recurrent conditioning but underutilize temporal priors. (b) Framework based on video DiT with 3D VAE decoder. Clip-wise methods capture richer spatiotemporal dependencies at the cost of multi-frame latency. (c) Our streaming framework based on video DiT.}
\label{fig:intro}
\end{figure}

Neural video compression (NVC) has achieved substantial progress by jointly optimizing conditional transform and entropy models, demonstrating competitive performance that surpasses traditional codecs~\cite{sullivan2012overview, bross2021overview, zhang2019recent}. Nevertheless, existing neural video codecs~\cite{li2024neural, jia2025towards, xiang2026real} remain limited at ultra-low bitrates, where distortion-oriented optimization tends to suppress uncertain high-frequency details, resulting in over-smoothed textures and degraded local structures. Consequently, enhancing the perceptual quality of reconstructions has become a central concern for extreme compression.

Motivated by recent advances in generative modeling, generative video compression (GVC) has emerged as a promising paradigm to address this challenge. Rather than explicitly preserving pixel-level details, GVC encodes key structural, motion, and semantic information into compact latent representations, enabling a generative decoder to synthesize plausible reconstructions. In particular, recent diffusion-based GVC methods~\cite{ma2025diffvcosd, xue2026single, li2026yoda, mao2026gnvcvd, xia2026vortec} leverage powerful generative priors to refine latent representations decoded by contextual neural codecs, greatly improving perceptual quality at ultra-low bitrates. Existing latent coding methods~\cite{ma2025diffvcosd, xia2026vortec, li2026yoda} rely on secondary transforms to directly compress generative latent features, while S$^2$VC~\cite{xue2026single} uses an adapter to map decoded features of a pixel codec into the diffusion latent space. Despite these different compression paradigms, the refined latent features still require multi-stage upsampling decoders for frame reconstruction, leading to additional computational overhead and decoding latency.

In addition, existing diffusion-based frameworks struggle to reconcile strong spatiotemporal modeling with causal frame-by-frame decoding. As illustrated in \cref{fig:intro}, frame-wise methods~\cite{ma2025diffvcosd, xue2026single, li2026yoda} employ image diffusion models for one-step refinement, conditioned on the temporal context derived from the previously reconstructed frame or contextual features. Although these approaches support online processing, their image diffusion backbones do not natively model spatiotemporal dependencies and require separate modules for temporal modeling. In contrast, clip-wise methods~\cite{mao2026gnvcvd, xia2026vortec} leverage video-native generative priors through joint refinement of multiple frames, inevitably introducing multi-frame buffering delay. Therefore, effectively leveraging video generative priors while preserving sequential decoding remains a challenge.

To address these limitations, we propose SGVC, a streaming generative video compression framework that adapts a pretrained latent video DiT into a causal pixel-space refiner. Rather than relying on the VAE latent space as an intermediate representation, SGVC directly maps decoded codec features to the input embedding space of the pretrained DiT. To bridge the representational gap between the codec and the DiT, we introduce Embedding Alignment during training, which aligns codec-derived embeddings with DiT input embeddings extracted from the original frames. The DiT then performs one-step refinement, and its final hidden states are mapped to the pixel domain through a lightweight reconstruction path, \textbf{bypassing the pretrained VAE decoder.}

To further exploit the spatiotemporal prior of the pretrained video DiT while maintaining streaming efficiency, we adapt its original bidirectional attention mechanism for causal refinement through In-Context Propagation, inspired by recent advances in autoregressive video generation and world models~\cite{yin2025slow, huang2026self, yang2026longlive}. Concretely, the I-frame refinement initializes the layer-wise K/V cache, providing temporal context for subsequent P-frames. During P-frame refinement, our In-Context Temporal Refinement blocks allow current-frame queries to attend to cached historical K/V states, which are subsequently updated for the next frame. By reusing the DiT's internal context, SGVC enables temporally coherent refinement without clip-wise buffering or repeated extraction of temporal guidance from previously refined reconstructions.

Extensive experiments demonstrate that SGVC achieves competitive rate--perception performance among generative video codecs supporting causal frame-by-frame decoding. Compared with S$^2$VC, SGVC achieves average BD-rate savings of 35.94 \ and 71.97\% in terms of LPIPS and DISTS, respectively. Notably, SGVC achieves the best FID-based coding efficiency among the compared methods, demonstrating a favorable trade-off between bitrate and reconstruction realism at ultra-low bitrates.

Our contributions are summarized as follows:
\begin{itemize}
    \item We propose SGVC, a streaming generative video compression framework that adapts a pretrained latent video DiT into a causal pixel-space refiner. SGVC directly reconstructs frames from refined DiT hidden states through a lightweight reconstruction path, bypassing the computationally expensive VAE decoder.
    \item We introduce an Embedding Alignment strategy during training that aligns codec-derived embeddings with  DiT input embeddings extracted from the original frames, bridging the representational gap between the codec and DiT.
    \item We develop an In-Context Propagation mechanism that adapts the pretrained bidirectional video DiT for causal frame-by-frame refinement. By propagating layer-wise K/V states across frames, SGVC exploits historical spatiotemporal context without clip-wise buffering or transmitting additional temporal side information.
\end{itemize}

\section{Related Work}

\subsection{Neural Video Compression for Fidelity}

Neural video compression typically adopts end-to-end optimized transform coding and entropy modeling to encode video frames into compact bitstreams. The DCVC series~\cite{li2023neural, li2024neural, li2026ultra} replaces the conventional residual-coding paradigm~\cite{lu2019dvc, feng2020learned} with contextual coding, in which each frame is encoded conditioned on temporal context derived from previously decoded references. This design substantially improves rate--distortion performance and has demonstrated competitiveness beyond traditional codecs. DCVC-RT~\cite{jia2025towards} extends this line of research by replacing explicit motion modeling with implicit feature propagation, thereby simplifying the coding architecture while outperforming VVC~\cite{bross2021overview} in terms of coding efficiency. More recently, UI$^2$C~\cite{xiang2026real} unifies intra- and inter-frame coding within the DCVC-RT framework, reducing the total parameter count. Collectively, recent studies have made substantial progress in rate-distortion performance and have increasingly focused on practical efficiency and model complexity. Nevertheless, their optimization objectives prioritize signal-level fidelity, making it difficult to preserve perceptual details at ultra-low bitrates. PLVC~\cite{yang2022plvc} instead improves perceptual quality through adversarial training, but its model capacity and training scale limit its performance at ultra-low bitrates.

\begin{figure*}[t]
\centering
\includegraphics[width=0.94\linewidth]{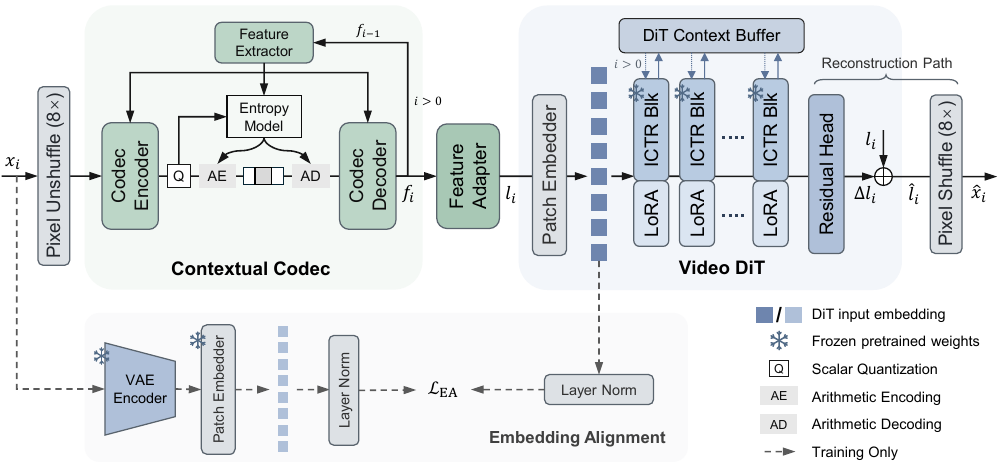}
\caption{\textbf{Overview of the proposed SGVC framework.} A contextual codec generates compact bitstreams, from which decoded features are projected into the DiT input embedding space for one-step refinement. The DiT’s layer-wise K/V states are initialized during I-frame refinement and propagated across subsequent P-frames, enabling causal frame-by-frame reconstruction without clip-wise buffering. 
A lightweight reconstruction path directly maps the refined DiT features to the pixel domain through residual prediction and pixel shuffle, bypassing the pretrained VAE decoder. During training, the lower branch performs Embedding Alignment by aligning codec-derived embeddings with DiT input embeddings extracted from the original frames.}
\label{fig:framework}
\end{figure*}

\subsection{Generative Video Compression}

To address the limitations of fidelity-oriented neural codecs at ultra-low bitrates, recent generative video compression (GVC) methods incorporate generative priors and explicitly optimize for perceptual quality. GLC \cite{jia2024generative, qi2025glc} performs transform coding in the latent space of VQGAN \cite{esser2021taming}, while GLVC \cite{guo2025glvc} combines intra- and inter-frame coding with a pretrained continuous tokenizer. GVC-1D~\cite{Zheng_2026_CVPR} further improves compression efficiency by employing compact 1D latent tokens and long-term memory to exploit spatial and temporal redundancy. Rather than directly compressing generative latent features, GVC-RT~\cite{dang2026gvcrt} employs a neural pixel codec to encode video frames and aligns decoded codec features with the latent representations of a pretrained LFQ tokenizer. Unlike this alignment for tokenizer-based reconstruction, our Embedding Alignment targets the DiT embedding space to facilitate generative refinement. Nevertheless, tokenizer-based approaches do not fully exploit the generative priors of pretrained diffusion models, motivating the exploration of diffusion-based GVC to further improve performance.

DiffVC~\cite{ma2025diffvc} employs an image diffusion model to refine the current frame conditioned on temporal context, while DiffVC-OSD~\cite{ma2025diffvcosd} and S$^2$VC \cite{xue2026single} further adapt one-step image diffusion models to perceptual video compression by injecting temporal or semantic context through dedicated adapters. YODA~\cite{li2026yoda} employs a one-step linear diffusion transformer (DiT) for generative refinement while incorporating multiscale temporal references into latent coding. To better capture spatiotemporal dependencies, pretrained video diffusion models (VDMs) have been adopted in GVC. M3-CVC~\cite{wan2025m3} and T-GVC~\cite{wang2026t} guide video generation using externally extracted text prompts and sparse trajectories, respectively. In contrast, GNVC-VD~\cite{mao2026gnvcvd} conditions a video DiT on decoded codec features for multi-step latent refinement, while VoRTeC~\cite{xia2026vortec} further introduces group-level one-step refinement. However, recent VDM-based methods~\cite{mao2026gnvcvd, xia2026vortec, xue2026generative, ma2026diffvcone} perform clip-level refinement by jointly processing multiple frames, leaving the causal adaptation of pretrained VDMs to streaming compression underexplored. Our method addresses this gap by refining each P-frame sequentially without multi-frame buffering.

\subsection{Streaming Generative Video Processing}

Recent studies have adapted pretrained video diffusion models for causal streaming generation. CausVid~\cite{yin2025slow} distills bidirectional video DiTs into few-step autoregressive models with K/V caching, while Self Forcing~\cite{huang2026self} mitigates exposure bias through autoregressive training rollouts. LongLive~\cite{yang2026longlive} further enables frame-level generation with short-window attention and frame sinks, while the Causal Forcing series~\cite{zhu2026causalforcing,zhao2026causalforcingpp} advances few-step causal distillation. For streaming video enhancement, FlashVSR~\cite{zhuang2026flashvsr} and SwiftVR~\cite{yan2026swiftvr} employ multi-frame processing with causal or limited-lookahead temporal dependencies to enable efficient inference. In contrast, SGVC reconstructs P-frames sequentially from decoded codec features without future-frame access, requiring adaptation to compression-induced degradations and propagation of the DiT's internal context.

\section{Methodology}

\subsection{Framework Overview}
\label{subsec:codec_feature}

\begin{figure}[t]
    \centering
    \includegraphics[width=0.96\linewidth]{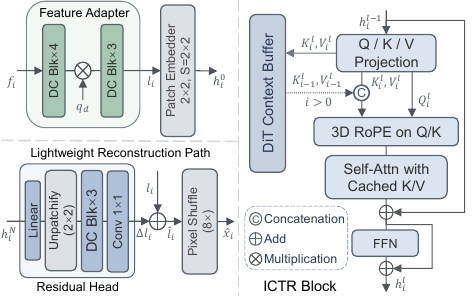}
    \caption{\textbf{Architecture of the feature adapter, ICTR block, and lightweight reconstruction path.} \emph{DC Blk} denotes the depthwise-separable convolutional block adopted from DCVC-RT. The scaling factor $q_d$ modulates intermediate features according to the compression level $q_i$. The ICTR block reuses layer-wise temporal context through cached K/V states, with 3D RoPE applied to queries and keys. Normalization layers are omitted for clarity.}
    \label{fig:ictr_block}
\end{figure}

To effectively exploit the spatiotemporal modeling capabilities of pretrained video diffusion models, we propose SGVC, a streaming generative video compression framework that combines a contextual pixel codec with a pretrained video Diffusion Transformer (DiT). As illustrated in \cref{fig:framework}, SGVC directly compresses each frame using the contextual pixel codec and adapts the decoded feature to the input embedding space of the DiT for one-step refinement. The refined DiT hidden states are then mapped directly to the pixel domain through a lightweight reconstruction path, bypassing the pretrained VAE decoder. \cref{fig:ictr_block} further provides a detailed view of the reconstruction branch, including the feature adapter, the In-Context Temporal Refinement (ICTR) block, and the lightweight reconstruction path.

Given the current frame $x_i$, the contextual codec first extracts temporal context $c_i$ from the propagated feature $f_{i-1}$ of the preceding frame. Conditioned on $c_i$ and compression level $q_i$, $x_i$ is spatially rearranged by an $8\times$ pixel-unshuffle operation and subsequently transformed into a latent feature $y_i$. The latent feature is then scalar-quantized and entropy-coded according to probability distributions estimated from temporal prior, hyperprior, and spatial context. At the decoder, the recovered feature is combined with the same temporal context $c_i$ to reconstruct the propagated feature $f_i$. This feature is causally reused to guide the coding of the next frame and simultaneously serves as a source-faithful anchor for generative refinement of the current frame. Following S$^2$VC~\cite{xue2026single} and GNVC-VD~\cite{mao2026gnvcvd}, we adopt the DCVC-RT \cite{jia2025towards} style architecture for contextual coding while eliminating the reliance on a pretrained VAE at the encoder.

\begin{figure}[t]
    \centering
    \includegraphics[width=0.95\linewidth]{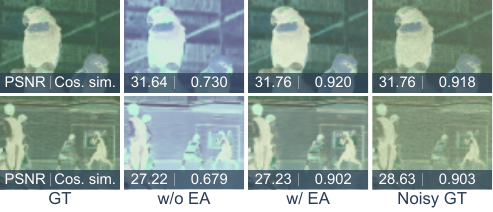}
    \caption{\textbf{PCA visualization of DiT input embeddings.}
    EA improves the alignment of codec-derived embeddings with the reference DiT embeddings. PSNR is computed against the ground-truth (GT) frames, while cosine similarity is measured against the GT embeddings. All embeddings are visualized using the same PCA projection and color normalization.}
    \label{fig:ea_pca}
\end{figure}

\subsection{Pixel-Space Refinement}
\label{subsec:pixel_refinement}
\label{subsec:one_step_refinement}

\paragraph{Codec Feature Adaptation}
To avoid relying on intermediate VAE latent representations during inference, the decoded codec feature $f_i$ is first mapped to a sub-pixel feature $l_i$ by a lightweight feature adapter $\mathcal{A}_{\mathrm{in}}$. A trainable patch embedder $\mathcal{P}_{\mathrm{in}}$ then projects this representation into the input embedding space of the pretrained DiT:
\begin{equation}
l_i=\mathcal{A}_{\mathrm{in}}(f_i, q_i),\qquad
h_i^{0}=\mathcal{P}_{\mathrm{in}}(l_i).
\label{eq:codec_feature_adaptation}
\end{equation}
However, the codec-derived embeddings may deviate from the input distribution expected by the pretrained DiT, potentially limiting the effectiveness of its generative priors.

\paragraph{Embedding Alignment}
To bridge the representational gap, we align the codec-derived embeddings with the DiT input embeddings obtained from the original frames during training. Specifically, a frozen teacher branch applies the pretrained VAE encoder $\mathcal{E}_{\mathrm{VAE}}$ and the original DiT patch embedder $\mathcal{P}_{\mathrm{pre}}$ to uncompressed frames:
\begin{equation}
h_i^{gt}
=\mathcal{P}_{\mathrm{pre}}\!\left(\mathcal{E}_{\mathrm{VAE}}(x_i)\right).
\label{eq:teacher_embedding}
\end{equation}
Each token of $h_i^{0}$ and $h_i^{gt}$ is normalized along the
embedding dimension, and the alignment loss is defined as
\begin{equation}
\mathcal{L}_{\mathrm{EA}}
=\mathbb{E}\left[
\left(
\operatorname{LN}_{D}(h_i^{0})
-\operatorname{sg}\left[\operatorname{LN}_{D}(h_i^{gt})\right]
\right)^2
\right],
\label{eq:embedding_alignment}
\end{equation}
where $\operatorname{LN}_{D}$ denotes token-wise layer normalization
without learnable affine parameters, and $\operatorname{sg}$
denotes the stop-gradient operation. As illustrated in \cref{fig:ea_pca}, EA increases the cosine similarity between codec-derived embeddings and the reference embeddings, indicating improved alignment with the pretrained DiT input space.

\paragraph{One-Step Refinement}
The pretrained video DiT learns a time-conditioned velocity field from a noise distribution toward the clean latent distribution under rectified flow. Rather than performing iterative sampling, we adapt the DiT for one-step residual refinement from compressed representations toward clean reconstructions. The ideal feature residual is given by $\tilde{l}_i-l_i$, where $\tilde{l}_i$ denotes the sub-pixel features obtained by applying $8\times$ pixel-unshuffle to $x_i$.

Given the codec-derived embeddings $h_i^0$, the adapted DiT
produces the final hidden states
$h_i^N$.
A lightweight residual head $\mathcal{R}_{\mathrm{out}}$
then estimates this residual:
\begin{equation}
\Delta l_i
=\mathcal{R}_{\mathrm{out}}(h_i^N),\quad
\hat{l}_i=l_i+\Delta l_i,\quad
\hat{x}_i=\mathcal{S}_{8}(\hat{l}_i),
\label{eq:feature_update}
\end{equation}
where $\Delta l_i$ denotes the predicted feature residual,
and $\mathcal{S}_{8}$ represents the $8\times$ pixel-shuffle operation.
The residual connection retains the decoded codec features
as content anchors while allowing the DiT to directly refine perceptual
details in the pixel domain.

\subsection{In-Context Propagation}
\label{subsec:temporal_propagation}

Effectively exploiting temporal correlations is essential for improving video compression efficiency. To adapt a pretrained bidirectional DiT to causal refinement, we equip each self-attention layer with a dedicated K/V cache, thereby converting each DiT block into an ICTR block. Following AEIC~\cite{zhang2026ultra}, we remove the text-conditioning cross-attention layers, as the clean codec-derived embeddings already encode semantic information. By propagating the temporal context encoded in the cached K/V states across consecutive frames, SGVC enables temporally coherent generative refinement under sequential frame reconstruction.

\paragraph{DiT Context Buffer}
In predictive video coding, each group of pictures (GOP) begins with an independently coded I-frame, followed by a sequence of predictively coded P-frames. Within each GOP, the ICTR blocks maintain and propagate layer-wise DiT context through a K/V buffer. For a DiT comprising $N$ ICTR blocks, the buffer after reconstructing frame $x_i$ is defined as
\begin{equation}
\mathcal{B}_i
=\left\{(K_i^l,V_i^l)\right\}_{l=1}^{N},
\label{eq:dit_context_buffer}
\end{equation}
where $K_i^l$ and $V_i^l$ denote the key and value states generated from frame $x_i$ at the $l$-th ICTR block. The buffer is initialized from the reconstructed I-frame and subsequently updated and propagated across P-frames, thereby providing in-context temporal conditioning throughout the GOP.

\paragraph{In-Context Initialization}
At the beginning of each GOP, the I-frame is decoded by the I-frame codec and refined by the video DiT without preceding temporal context. During the refinement, each ICTR block computes the I-frame key and value states:
\begin{equation}
\left(K_0^l,V_0^l\right)
=
\left(
W_K^l h_0^{l-1},\,
W_V^l h_0^{l-1}
\right),
\quad l=1,\ldots,N,
\label{eq:iframe_kv}
\end{equation}
where $h_0^{l-1}$ denotes the I-frame hidden state before the $\ell$-th ICTR block, while $W_K^l$ and $W_V^l$ are the corresponding key and value projection matrices. These states initialize the DiT context buffer $\mathcal{B}_0$, allowing the queries of the first P-frame to attend directly to the cached I-frame context within the inherent self-attention operation, without introducing an additional temporal-conditioning branch.

\paragraph{Cross-Frame K/V Propagation}
During the reconstruction of P-frame $\hat{x}_i$ ($i>0$), the hidden states at the $l$-th ICTR block are projected into the current queries, keys, and values, denoted by $Q_i^l$, $K_i^l$, and $V_i^l$, respectively. The current-frame K/V states are concatenated with the cached K/V states from the context buffer. Before self-attention, 3D RoPE is applied to $Q_i^l$ and the concatenated keys according to their temporal position and spatial coordinates associated with each token. The current queries then attend to the concatenated key--value pairs. For notational simplicity, the RoPE operations are omitted from the following equations:
\begin{align}
\widetilde{K}_i^l
&=
\left[K_{i-1}^l;K_i^l\right],
\quad
\widetilde{V}_i^l
=
\left[V_{i-1}^l;V_i^l\right],
\label{eq:cached_kv}
\\
A_i^l
&=
\operatorname{softmax}\!\left(
\frac{
Q_i^l
\left(\widetilde{K}_i^l\right)^\top
}{
\sqrt{d}
}
\right)
\widetilde{V}_i^l.
\label{eq:cached_attention}
\end{align}
Here, $[\cdot;\cdot]$ denotes concatenation along the token dimension, and $d$ is the dimension of each attention head. After reconstructing $\hat{x}_i$, the context buffer is updated by discarding the previous states and retaining only $(K_i^l,V_i^l)$ at each ICTR block. The buffer therefore maintains a constant memory footprint and propagates temporal information implicitly.

\subsection{Training Strategy}
\label{subsec:training}

We adopt a two-stage training strategy to progressively incorporate the generative prior into the compression framework.

\paragraph{Stage 1: Codec Pretraining}
We first optimize the contextual pixel codec and adaptation modules without DiT refinement. The objective combines rate regularization, auxiliary reconstruction, and embedding alignment:
\begin{equation}
\mathcal{L}_{\mathrm{codec}}
= \mathcal{R} + \lambda_r
\left(
\mathcal{L}_{\mathrm{aux}}
+0.1\mathcal{L}_{\mathrm{EA}}
\right),
\label{eq:codec_training}
\end{equation}
where $\mathcal{R}$ denotes the estimated bits per pixel and Lagrange multiplier $ \lambda_r$ balances the rate-distortion trade-off. The auxiliary reconstruction loss is defined as
\begin{equation}
\mathcal{L}_{\mathrm{aux}}
= \mathcal{L}_{\mathrm{MSE}}(x_i,\bar{x}_i)
+ 0.1\mathcal{L}_{\mathrm{LPIPS}}(x_i,\bar{x}_i),
\label{eq:auxiliary_loss}
\end{equation}
where $\bar{x}_i$ denotes the auxiliary reconstruction obtained by applying $8\times$ pixel-shuffle to the sub-pixel feature $l_i$ produced by the feature adapter.

\paragraph{Stage 2: Joint Generative Training}
We then introduce the pretrained video DiT and jointly optimize the codec, adaptation modules, reconstruction path, and LoRA parameters, while keeping the pretrained DiT backbone and alignment teacher frozen. The training objective becomes
\begin{equation}
\begin{aligned}
\mathcal{L}_{\mathrm{joint}}
= \mathcal{R} + \lambda_r \bigl(
&\mathcal{L}_{\mathrm{rec}}
+ 0.1\mathcal{L}_{\mathrm{aux}}
+0.1\mathcal{L}_{\mathrm{EA}}
\bigr),
\end{aligned}
\label{eq:joint_training}
\end{equation}
where the reconstruction loss combines fidelity, perceptual, and adversarial supervision:
\begin{equation}
\begin{aligned}
\mathcal{L}_{\mathrm{rec}}
={}&\mathcal{L}_{\mathrm{MSE}}(x_i,\hat{x}_i)
+\mathcal{L}_{\mathrm{LPIPS}}(x_i,\hat{x}_i)\\
&+\mathcal{L}_{\mathrm{DISTS}}(x_i,\hat{x}_i)
+0.1\mathcal{L}_{\mathrm{adv}}.
\end{aligned}
\label{eq:reconstruction_loss}
\end{equation}
Following~\cite{zhang2026ultra}, we employ LPIPS-VGG~\cite{zhang2018unreasonable} and overlap-chunked edge-aware DISTS~\cite{li2026unleashing} for perceptual supervision, together with a DINOv2-based~\cite{oquab2023dinov2} adversarial loss using a frozen backbone and a trainable discriminator head.

\begin{figure*}[t]
    \centering
    \includegraphics[width=0.97\linewidth]{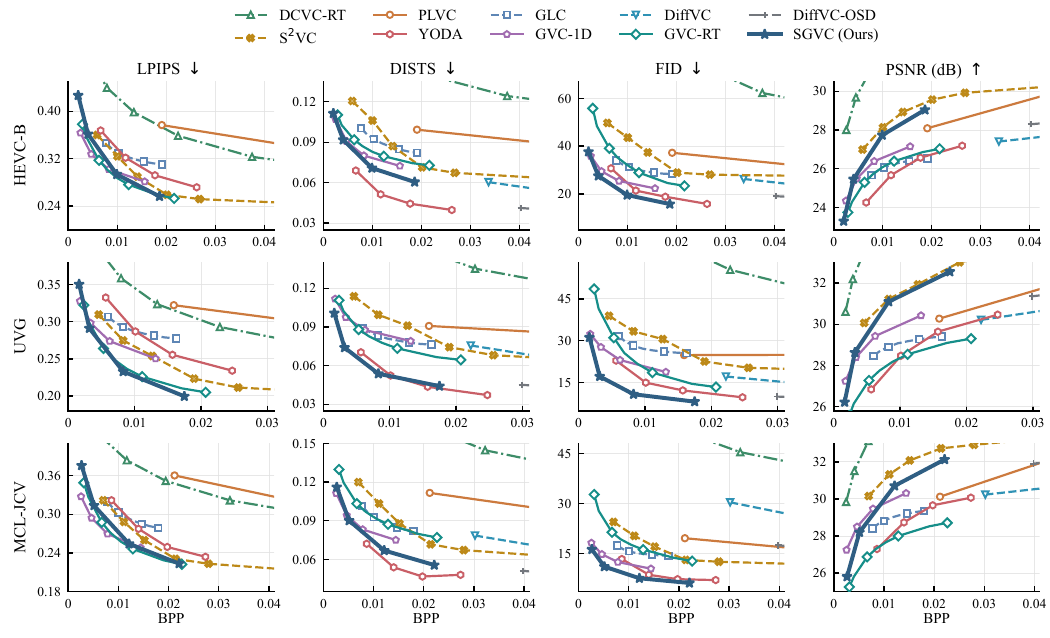}
    \caption{Rate--distortion--perception comparisons with existing video compression methods on HEVC Class B, UVG, and MCL-JCV.}
    \label{fig:rd_main}
\end{figure*}

\relaxexperimentspacing
\section{Experiments}

\subsection{Experimental Setup}
\noindent\textbf{Implementation Details.}
SGVC is trained on Vimeo-90K~\cite{xue2019video} using $256\times256$ crops. The DiT is initialized from Wan-2.1-T2V-1.3B~\cite{wan2025wan} and adapted using LoRA~\cite{hu2022lora} with a rank of 64. To support multiple bitrates with a single model, we randomly sample one of four compression levels at each iteration, with corresponding $\lambda_r \in \{0.05, 0.10, 0.25, 0.50\}$. Following DCVC-DC~\cite{li2023neural}, we adopt hierarchical distortion weights to support a hierarchical quality structure. We progressively increase the training sequence length with a decaying learning rate and use a global batch size of 32 across two NVIDIA H100 GPUs.

For inference, we adopt spatially local attention implemented with FlexAttention~\cite{dong2025flexattention} to reduce computational overhead and mitigate the resolution gap between training and inference, with further details provided in the appendix. 

\begin{figure}[t]
    \raggedright
    \includegraphics[width=0.96\linewidth]{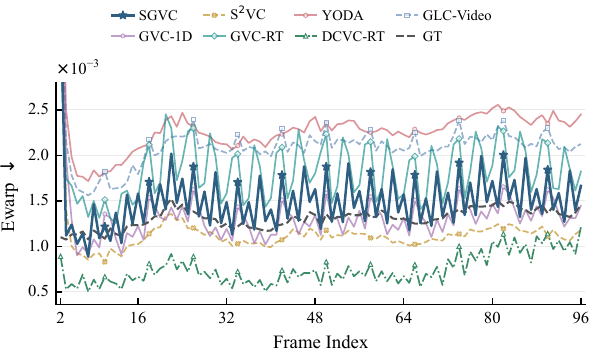}
    \caption{
    \textbf{Per-frame temporal consistency comparison on UVG at comparable bitrates.}
    The corresponding bitrates are 0.0066 (SGVC), 0.0068 (S$^2$VC), 0.0093 (YODA), 0.0055 (GVC-1D), 0.0078 (GVC-RT), 0.0076 (GLC), and 0.0057 (DCVC-RT) bpp.
    }
    \label{fig:ewarp_by_frame}
\end{figure}

\begin{figure*}[t]
    \centering
    \includegraphics[width=0.98\textwidth]{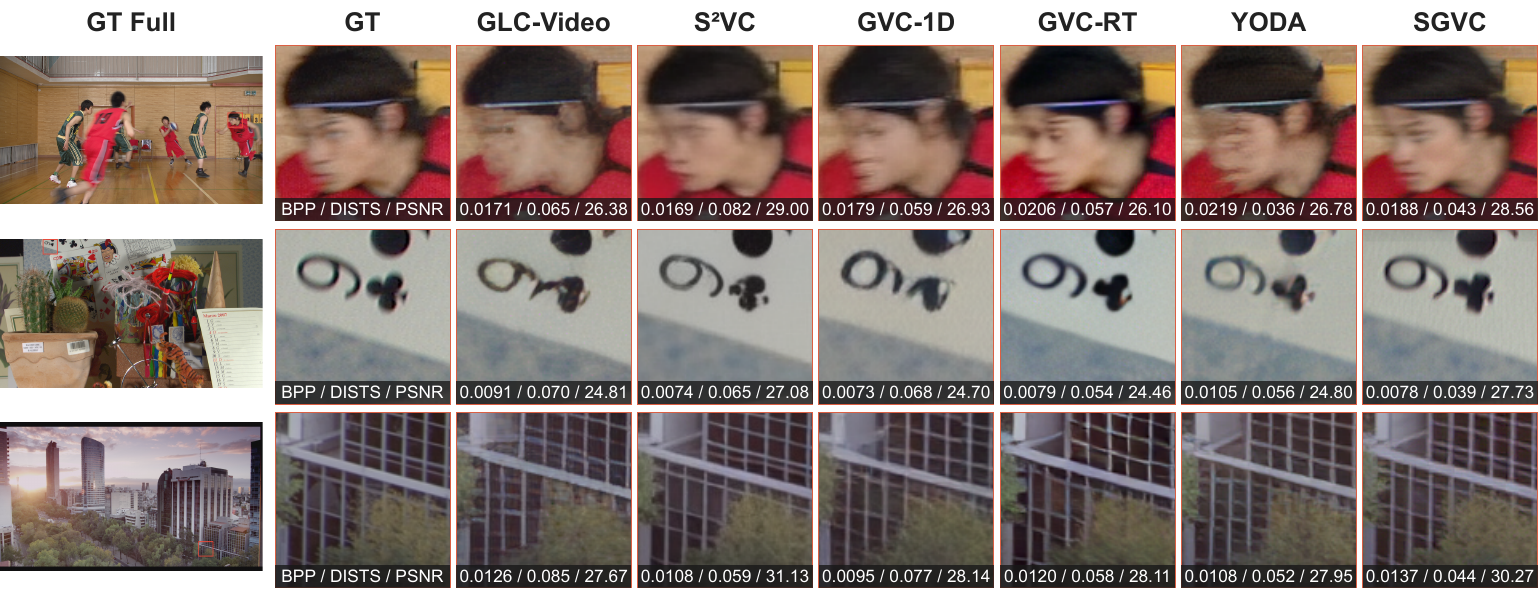}
    \caption{Qualitative comparison between ground-truth (GT) and
reconstructed frames produced by different generative video codecs.}
    \label{fig:qualitative}
\end{figure*}

\begin{figure}[t]
    \centering
    \includegraphics[width=\linewidth]{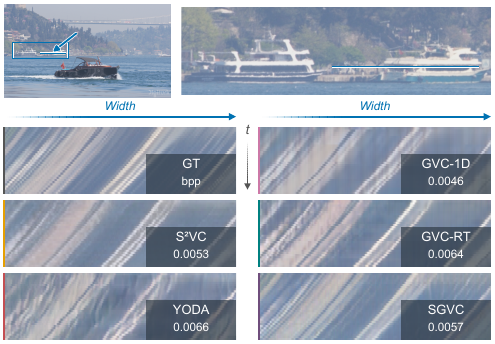}
    \caption{\textbf{Visual comparison of temporal consistency.}
    The spatiotemporal slices are constructed by stacking the regions indicated by the blue lines across consecutive frames along the temporal axis.}
    \label{fig:temporal_slice}
\end{figure}

\paragraph{Evaluation Settings}
We compare SGVC with learned video codecs under the low-delay P (LDP) configuration, including DCVC-RT~\cite{jia2025towards}, PLVC~\cite{yang2022plvc}, GLC~\cite{qi2025glc}, GVC-RT~\cite{dang2026gvcrt}, GVC-1D~\cite{Zheng_2026_CVPR}, DiffVC~\cite{ma2025diffvc}, DiffVC-OSD~\cite{ma2025diffvcosd}, S$^2$VC~\cite{xue2026single}, and YODA~\cite{li2026yoda}. Since DiffVC and DiffVC-OSD are not publicly available, we use the results reported in their original papers. PLVC is evaluated on an NVIDIA GeForce RTX 2080 Ti GPU, while the remaining codecs are evaluated on a single NVIDIA GeForce RTX 3090 GPU.

Experiments are conducted on HEVC Class B~\cite{bossen2010common}, MCL-JCV~\cite{wang2016mcl}, and UVG~\cite{mercat2020uvg}. Following DiffVC, we evaluate the first 96 frames of each sequence with an intra period of 32. Bitrate is measured in bits per pixel (BPP), 
while reconstruction quality is evaluated using PSNR for signal fidelity, LPIPS~\cite{zhang2018unreasonable} and DISTS~\cite{ding2022image} for perceptual quality, as well as FID~\cite{heusel2017gans} for distributional similarity. Following GVC-RT~\cite{dang2026gvcrt}, we adopt VGG-based LPIPS for its better alignment with subjective perceptual quality in generative compression. To further evaluate temporal consistency, we measure the RAFT-based warping error $E_{\mathrm{warp}}$~\cite{lai2018learning,teed2020raft} over the first 96 frames of each sequence in UVG at comparable bitrates, with the intra period set to 96. Optical flow is estimated from ground-truth frames, with occluded and out-of-bound regions excluded from evaluation.

\subsection{Quantitative Comparison}
\cref{fig:rd_main} shows the rate--distortion--perception performance of SGVC and existing video compression methods that support frame-by-frame compression. SGVC achieves favorable perceptual quality at ultra-low bitrates ($<0.03$ bpp), particularly in terms of DISTS and FID. Compared with S$^2$VC, SGVC reduces the average BD-rate~\cite{bjontegaard2001calculation} by $35.94\%$, $71.97\%$, and $89.15\%$ in LPIPS, DISTS, and FID, respectively, across the three datasets. SGVC also attains the lowest BD-rate in FID among the compared methods.

Compared with GVC-RT, SGVC achieves lower BD-rates in DISTS and FID across all three datasets, while also outperforming GVC-1D in FID. SGVC outperforms YODA in LPIPS and FID, while YODA retains an advantage in DISTS on average. On UVG, SGVC achieves the lowest BD-rates of $-50.96\%$ and $-85.41\%$ in LPIPS and DISTS. Furthermore, the PSNR curves show that SGVC generally preserves higher signal fidelity than GVC-RT, GVC-1D, and YODA.

\subsection{Qualitative Comparison}

\cref{fig:qualitative} presents a qualitative comparison across different generative video codecs. S$^2$VC preserves the overall structure of the original frames but loses fine-grained details, resulting in smoother reconstruction. GVC-RT prioritizes computational efficiency, while its reconstructions exhibit fewer details and noticeable color shifts. In contrast, methods that directly compress generative latent representations, such as GLC, GVC-1D, and YODA, produce sharper details, yet their reconstructed textures and structures often deviate from the original content. SGVC preserves rich texture details while producing more realistic reconstructions that better match the original frames, consistent with the quantitative analysis.

\cref{fig:ewarp_by_frame} and \cref{fig:temporal_slice} further illustrate temporal consistency through per-frame warping errors and spatiotemporal slices. SGVC achieves lower warping errors than YODA and GVC-RT over most frame positions, while the reconstructions of YODA and GVC-RT exhibit irregular texture variations across frames in \cref{fig:temporal_slice}. S$^2$VC exhibits smoother temporal variations, with warping errors even lower than those of the ground truth (GT) at some frame positions, but its reconstructions tend to lose fine texture details. GVC-1D maintains strong temporal continuity, potentially benefiting from its long- and short-term temporal modeling, although its reconstructed textures appear less realistic. SGVC exhibits coherent temporal variations that closely follow the original sequence while preserving rich texture details.

\begin{table}[t]
\centering
\footnotesize
\setlength{\tabcolsep}{3pt}
\renewcommand{\arraystretch}{1.08}

\caption{\textbf{I-frame ablation on DIV2K.}
BD-rate (\%, $\downarrow$) is evaluated in terms of PSNR, LPIPS, DISTS, and FID, with SGVC as the anchor. Decoding time is measured on DIV2K images resized to 1080p using an NVIDIA RTX 3090 GPU. EA and LA denote embedding alignment and latent alignment, respectively.}

\label{tab:ablation_intra}
\begin{tabular*}{\linewidth}{
@{\extracolsep{\fill}}lrrrrr@{}}
\toprule
Variant & PSNR & LPIPS & DISTS & FID & Dec. (ms) \\
\midrule
\multicolumn{6}{@{}l}{
\textit{I. Generative Refinement}} \\
w/o DiT (aux.)
& \textbf{-17.45} & 76.88 & 263.36 & 292.49 & 198.18 \\
w/o DiT (retrained)
& 2.54 & 27.34 & 41.80 & 93.24 & 197.29 \\
w/o EA
& -2.05 & 26.62 & 42.40 & 99.58 & 889.88 \\
\textbf{SGVC (ours)}
& 0.00 & \textbf{0.00} & \textbf{0.00} & \textbf{0.00} & 891.36 \\
\midrule
\multicolumn{6}{@{}l}{
\textit{II. Compression Pipeline}} \\
Latent coding
& 26.01 & 138.49 & 298.00 & 280.70 & 1538.44 \\
LA + VAE dec.
& \textbf{-7.60} & \textbf{-10.87} & \textbf{-4.23} & \textbf{-18.75} & 1498.54 \\
\textbf{SGVC (ours)}
& 0.00 & 0.00 & 0.00 & 0.00 & \textbf{891.36} \\
\bottomrule
\end{tabular*}
\end{table}

\subsection{Ablation Study}

We conduct ablation studies to examine the contributions of SGVC's key components to perceptual reconstruction and decoding efficiency. We first investigate generative refinement and alternative compression pipelines on I-frames, followed by an evaluation of embedding alignment (EA) and in-context propagation (ICP) on P-frames.

\paragraph{Generative Refinement}
\Cref{tab:ablation_intra} evaluates the contributions of the pretrained video DiT and Embedding Alignment (EA) to I-frame reconstruction. We consider two variants without DiT refinement: \emph{w/o DiT (aux.)} directly uses the auxiliary reconstruction output, while \emph{w/o DiT (retrained)} retrains the framework without the DiT. Even after retraining, removing the DiT increases BD-rate by 27.34\%, 41.80\%, and 93.24\% in terms of LPIPS, DISTS, and FID, respectively. These results demonstrate that the pretrained video DiT substantially improves perceptual reconstruction quality and visual realism. The variant \emph{w/o EA} removes the embedding alignment objective while retaining SGVC, resulting in BD-rate increases of 26.62\%, 42.40\%, and 99.58\% in the three metrics. EA provides an additional supervision signal during training, encouraging codec-derived embeddings to be more compatible with the DiT for pixel-space refinement.

\paragraph{Compression Pipeline}
We further evaluate two alternative compression pipelines. \emph{Latent coding} directly compresses VAE latent features instead of the $8\times$ pixel-unshuffled features used by SGVC, while \emph{LA + VAE dec.} aligns decoded codec features with the pretrained VAE latent space and reconstructs frames using the VAE decoder rather than the reconstruction head. Direct latent coding incurs higher BD-rates across all four metrics while increasing decoding time from 891.36 to 1538.44 ms. The variant \emph{LA + VAE dec.} achieves better rate--distortion--perception performance but requires 1498.54 ms for decoding. By bypassing the VAE decoder, SGVC reduces decoding time by 40.52\% at the cost of a moderate decrease in reconstruction quality.

\begin{figure}[t]
    \centering
    \includegraphics[width=\linewidth]{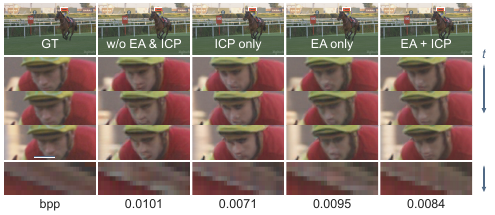}

    \caption{\textbf{Qualitative ablation of EA and ICP.}
    The top row shows reconstructed frames, while the lower rows illustrate temporal variations through consecutive frames and spatiotemporal slices.}

    \label{fig:ablation_visual}
\end{figure}

\paragraph{Embedding Alignment and In-Context Propagation}
To further validate the two key components of SGVC, we evaluate the effects of EA and ICP on P-frame reconstruction while keeping the I-frame model frozen. We use FloLPIPS~\cite{danier2022flolpips} to jointly assess perceptual quality and temporal consistency. As shown in \Cref{tab:ablation_inter}, removing both components increases BD-rate by 51.93\% and 44.49\% in DISTS and FloLPIPS, respectively. Introducing either component improves reconstruction quality, while their combination achieves the best performance in both metrics. With EA enabled, ICP increases decoding time and peak GPU memory but further improves temporal consistency, as reflected by BD-rate in FloLPIPS. \cref{fig:ablation_visual} further shows that the full model better preserves facial details and temporal continuity.

\begin{table}[t]
\centering
\small
\setlength{\tabcolsep}{3pt}
\renewcommand{\arraystretch}{1.09}

\caption{\textbf{P-frame ablation on UVG.}
BD-rate (\%, $\downarrow$) is evaluated in DISTS and FloLPIPS, with the complete SGVC (EA + ICP) as the anchor. Decoding time and peak GPU memory are also reported.}
\label{tab:ablation_inter}

\begin{tabular*}{\linewidth}{@{\extracolsep{\fill}}ccrrrr@{}}
\toprule
EA & ICP & DISTS & FloLPIPS & Dec. (ms) & Mem. (GiB) \\
\midrule
$\times$ & $\times$ & 51.93 & 44.49 & 727.44 & 4.51 \\
$\times$ & $\checkmark$ & 8.08 & 17.08 & 866.99 & 5.94 \\
$\checkmark$ & $\times$ & 0.11 & 16.45 & 728.37 & 4.51 \\
\midrule
$\checkmark$ & $\checkmark$ & \textbf{0.00} & \textbf{0.00} & 867.58 & 5.94 \\
\bottomrule
\end{tabular*}
\end{table}

\vspace*{2.618bp}
\section{Conclusion}
We presented SGVC, a streaming generative video compression framework that adapts a pretrained latent video DiT into a causal pixel-space refiner. SGVC directly projects decoded codec features into the DiT input embedding space, with Embedding Alignment bridging the representation gap. In-Context Propagation reuses layer-wise K/V states across frames to exploit historical generative context without clip-wise buffering, while a lightweight reconstruction path maps the refined features directly to the pixel domain, bypassing the pretrained VAE decoder to reduce decoding overhead.

\clearpage
\restorefloatspacing
{\small
\bibliographystyle{ieeenat_fullname}
\bibliography{main}
}

\clearpage
\appendix
\twocolumn[{\centering\Large\bfseries
Supplementary Material for\\
Streaming Generative Video Compression\\
with Embedding Alignment and In-Context Propagation\par\vspace{18pt}}]

\section{Additional Implementation Details}
\label{sec:supp_implementation}

\subsection{Training Details}

\begin{figure}[t]
    \centering
    \includegraphics[width=\linewidth]{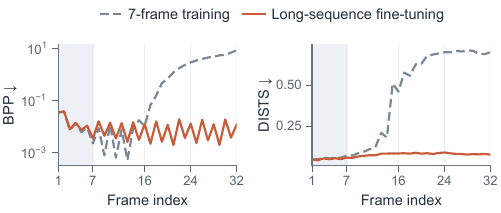}
    \caption{\textbf{Effect of long-sequence fine-tuning on
    per-frame BPP and DISTS.}
    Fine-tuning substantially
    mitigates bitrate growth and reconstruction degradation
    over extended sequences.}
    \label{fig:supp_longseq_ft}
\end{figure}

We use AdamW~\cite{loshchilov2019decoupled} to train I-frame
and P-frame models independently.
The initial learning rates are $10^{-5}$ for the codec and
$10^{-4}$ for the remaining trainable modules and
discriminator. Both follow a cosine decay
schedule, with a minimum
learning rate ratio of 0.05. We use two NVIDIA
H100 GPUs, with a batch size of 4 per GPU and
4 gradient accumulation steps, yielding a
global batch size of 32.

\paragraph{I-Frame Training}
We first train the I-frame model for 90,000 optimization steps.
During the first 2,000 steps, we optimize the codec and
adaptation modules using the auxiliary reconstruction loss,
without DiT refinement. Embedding Alignment (EA) is enabled
from the beginning to supervise the trainable patch embedder,
while the teacher patch embedder remains frozen throughout
training. We then introduce the pretrained video DiT and
jointly optimize the codec, adaptation modules, DiT LoRA
parameters, and residual head. The final reconstruction loss
is introduced at this stage, followed by adversarial training
after 40,000 steps.

\paragraph{P-Frame Training}
We train the P-frame model for 30,000 steps while
keeping the I-frame model frozen. During the first
1,000 steps, we optimize the contextual codec and
adaptation modules without DiT refinement. We then
introduce the pretrained video DiT for joint training,
with adversarial training enabled after 20,000 steps.

We adopt a progressive training
strategy. Specifically, we train on two-frame
sequences for the
first 5,000 steps, including the initial I-frame.
The sequence length then increases
by one frame every 1,000 steps, reaching seven frames
at step 9,000 and remaining unchanged for the rest of training.
Following the cascaded training strategy of DCVC-DC~\cite{li2023neural},
we accumulate gradients over successive P-frames and
update model parameters only after completing
gradient accumulation across sequences.

\begin{figure}[t]
    \centering
    \includegraphics[width=\linewidth]{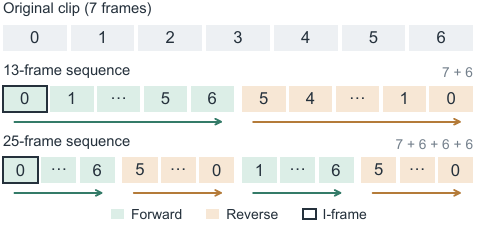}
    \caption{\textbf{Construction of extended training sequences
    through temporal reversal and concatenation.}
    The original 7-frame clip is extended into 13-frame and
    25-frame sequences.}
    \label{fig:supp_seq_construction}
\end{figure}

\begin{figure}[t]
    \centering
    \includegraphics[width=\linewidth]{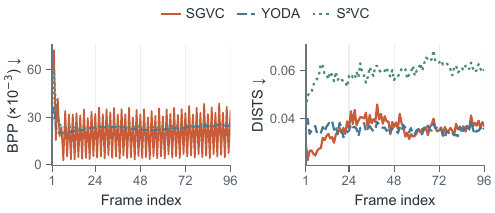}
    \caption{\textbf{Long-sequence evaluation with an intra
    period of $-1$}. Per-frame BPP and DISTS are compared
    across SGVC, YODA~\cite{li2026yoda}, and S$^2$VC~\cite{xue2026single}.}
    \label{fig:supp_longseq_perframe}
\end{figure}

\begin{figure*}[t]
    \centering
    \includegraphics[width=\textwidth]{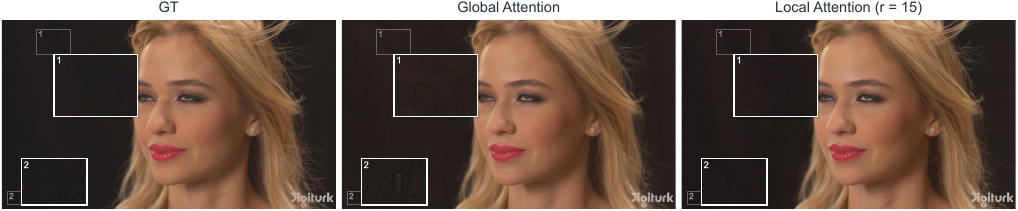}
    \caption{\textbf{Visual comparison of global and
    spatially local attention on the \emph{Beauty}
    sequence from UVG.}
    At $1920\times1080$, spatially local attention
    ($r=15$) mitigates reconstruction artifacts
    in smooth regions.}
    \label{fig:supp_local_visual}
\end{figure*}

\begin{figure*}[t]
    \centering
    \includegraphics[width=\textwidth]{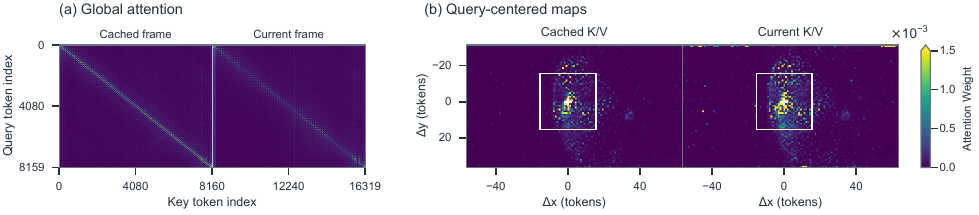}
    \caption{\textbf{Visualization of global attention weights.}
    Attention weights are jointly normalized over cached
    and current-frame keys and averaged across 12 heads.
    (a) Block-averaged query--key attention matrix.
    (b) Spatial attention maps for the same query.}
    \label{fig:supp_attention}
\end{figure*}

\begin{table}[t]
\centering
\footnotesize
\setlength{\tabcolsep}{3pt}
\renewcommand{\arraystretch}{1.08}

\caption{\textbf{Comparison of global and spatially
local attention.}
Results are averaged over 32 frames from each of
the \emph{Beauty}, \emph{Jockey}, and \emph{ShakeNDry}
sequences in UVG.
Decoding time is measured per P-frame, and memory
denotes peak allocated GPU memory.}

\label{tab:supp_local_attention}

\begin{tabular*}{\linewidth}{
@{\extracolsep{\fill}}lrrrr@{}}
\toprule
Attention & PSNR $\uparrow$ & DISTS $\downarrow$
& Dec. (ms) $\downarrow$ & Mem. (GiB) $\downarrow$ \\
\midrule
\multicolumn{5}{@{}l}{
\textit{$512\times512$}} \\
Global & \textbf{30.35} & 0.1111 & 102.77 & 2.615 \\
Local ($r=15$) & 30.33 & \textbf{0.1064}
& \textbf{101.25} & 2.615 \\
\midrule
\multicolumn{5}{@{}l}{
\textit{$1920\times1080$}} \\
Global & 31.04 & 0.0945 & 1134.63 & 5.494 \\
Local ($r=15$) & \textbf{31.43} & \textbf{0.0669}
& \textbf{795.85} & 5.494 \\
\bottomrule
\end{tabular*}
\end{table}

\paragraph{Long-Sequence Fine-Tuning}

Following common practice in neural video compression,
we train SGVC on Vimeo-90K~\cite{xue2019video},
which provides only seven consecutive frames per clip.
As shown in~\cref{fig:supp_longseq_ft}, the model trained on sequences
of up to seven frames suffers from severe error accumulation
when evaluated on longer sequences, leading to a rapid
increase in bitrate and perceptual distortion.
To mitigate this issue, we further fine-tune the P-frame
model for 4,000 steps using extended training sequences.
As illustrated in \cref{fig:supp_seq_construction},
we construct 13-frame and 25-frame sequences through
temporal reversal and concatenation of the original
seven-frame clips. The first 1,000 steps use 13 frames,
followed by 3,000 steps with 25 frames.

To account for artificial motion reversals and reduce
memory consumption, we adopt a group-wise training strategy
inspired by ECVC~\cite{jiang2025ecvc}, with group
boundaries aligned with the motion reversal points.
We perform backpropagation after each group and
detach the recurrent codec features and DiT K/V
states before processing the next group.
Gradients are accumulated across groups, while
model parameters are updated after processing
the complete sequence. This limits gradient propagation across artificial
motion reversal boundaries while maintaining
longer prediction chains.

As shown in \cref{fig:supp_longseq_ft}, long-sequence
fine-tuning substantially mitigates bitrate growth
and perceptual degradation. Further evaluation
without periodic intra refresh in
\cref{fig:supp_longseq_perframe} shows that SGVC maintains
stable bitrate and competitive perceptual quality
over 96 frames. Nevertheless, some degradation
in perceptual quality persists, which may be
further mitigated by fine-tuning on genuine
long video sequences.

\subsection{Spatially Local Attention}
\label{sec:supp_local_attention}

SGVC is trained on $256\times256$ crops, resulting
in a resolution mismatch during high-resolution
inference. As shown in~\cref{fig:supp_local_visual},
global attention introduces noticeable artifacts
in smooth regions.
\cref{fig:supp_attention} further reveals that attention
weights are largely concentrated on spatially
neighboring tokens.

We therefore adopt spatially local attention
implemented with FlexAttention~\cite{dong2025flexattention}.
Specifically, each query attends to a $31\times31$
square window ($r=15$) in the DiT token grid.
The same spatial constraint applies to both
current-frame and cached previous-frame K/V states,
preserving causal temporal conditioning.

As shown in~\cref{tab:supp_local_attention},
with identical bitstreams, global and local attention
achieve comparable performance at $512\times512$.
At $1920\times1080$, local attention achieves
more substantial improvements, increasing PSNR
by 0.39 dB, reducing DISTS by 29.2\%, and
decreasing decoding time by 29.9\%.

\begin{figure*}[t]
    \centering
    \includegraphics[width=\textwidth]{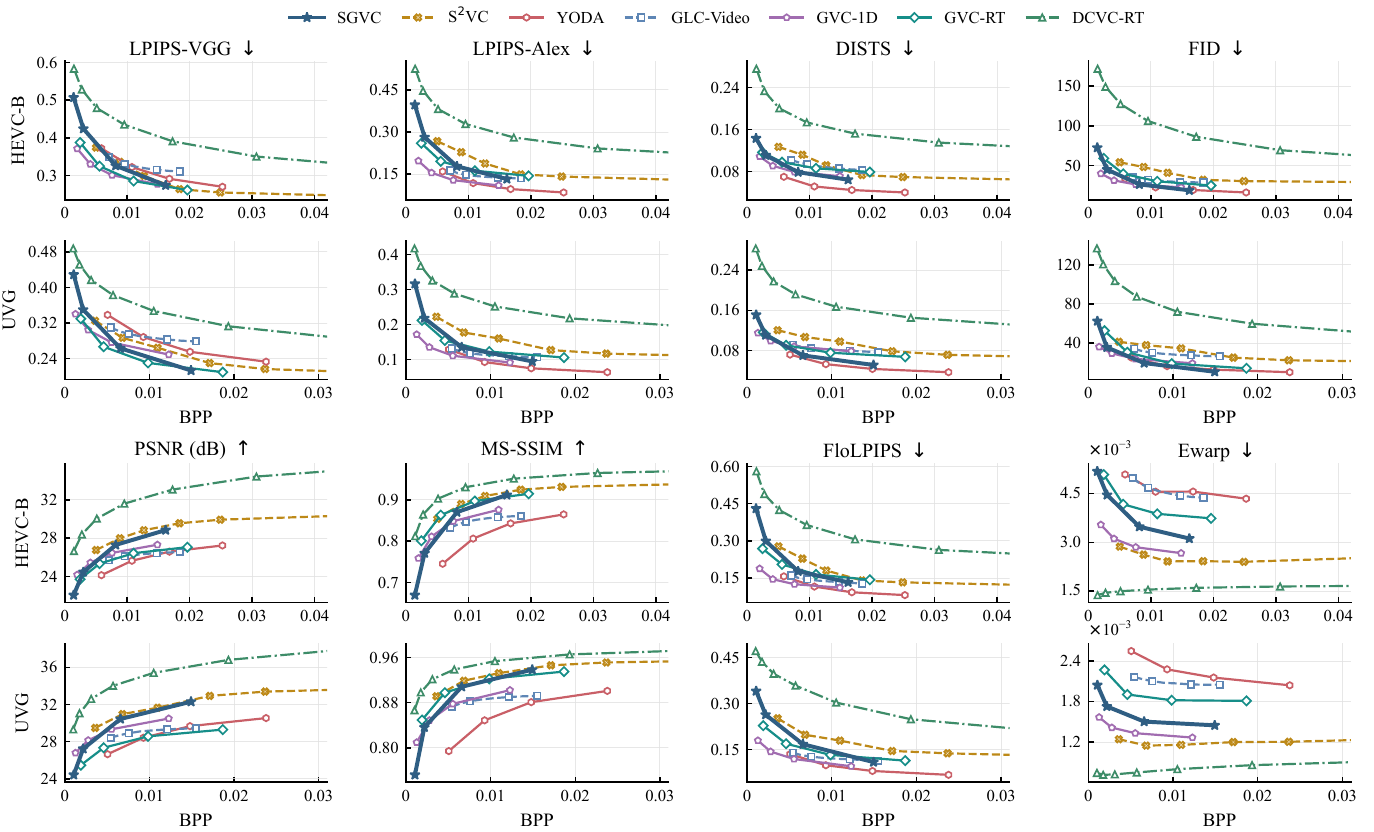}
    \caption{\textbf{Rate--distortion--perception curves
    on HEVC Class B and UVG.}
    The first 96 frames of each sequence are evaluated
    with a single intra frame (intra-period = $-1$).}
    \label{fig:supp_longseq_rd}
\end{figure*}

\begin{figure}[t]
    \centering
    \includegraphics[width=1.0\linewidth]{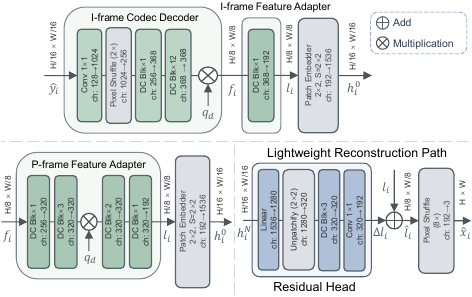}
    \caption{\textbf{Detailed network architectures.}
    DC Blk $\times n$ denotes $n$ consecutive
    depthwise-separable convolutional blocks.
    Channel dimensions are indicated for each module.
    The I-frame and P-frame models use the same
    lightweight reconstruction architecture.}
    \label{fig:supp_network}
\end{figure}

\subsection{Network Structure}
\label{sec:supp_network}

SGVC builds on the contextual coding architecture
of DCVC-RT~\cite{jia2025towards}, while focusing
on producing decoded representations better suited
for generative refinement with a pretrained video DiT.
 Given the higher complexity of the I-frame codec decoder, we simplify its feature adapter to a DC block adopted from DCVC-RT~\cite{jia2025towards}.
\Cref{fig:supp_network} provides detailed architectures
of the I-frame codec decoder, I-frame and P-frame
feature adapters, and lightweight reconstruction path,
including channel dimensions.

\begin{figure*}[t]
    \centering
    \includegraphics[width=0.96\textwidth]{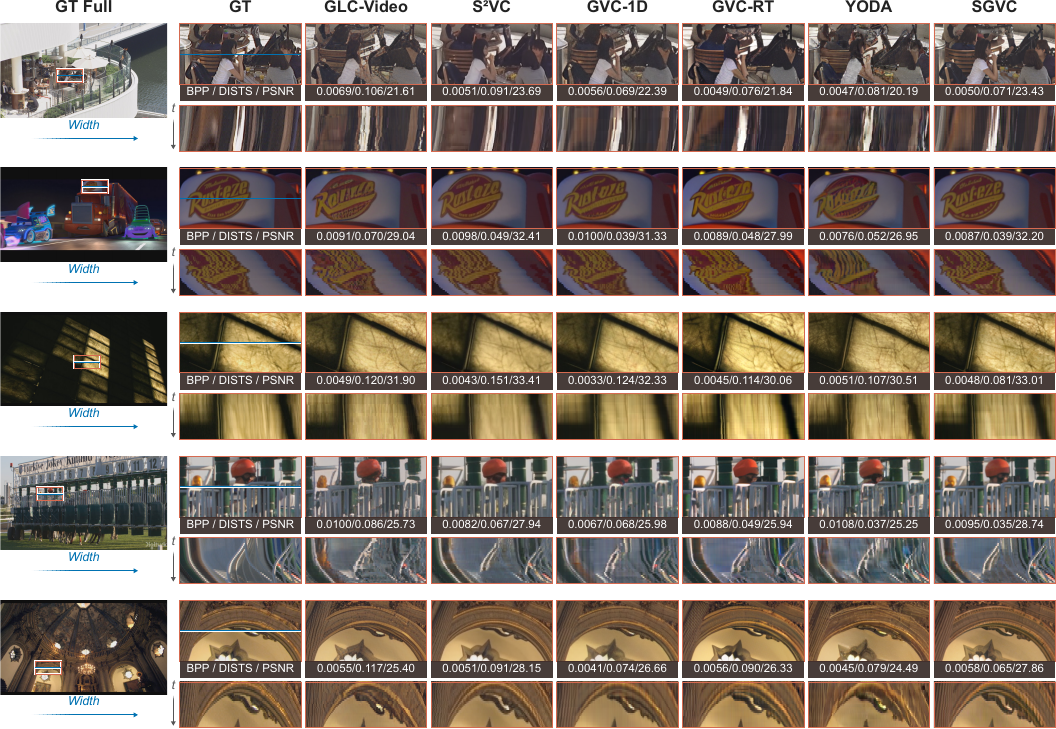}
    \caption{\textbf{Qualitative comparison of spatial and temporal reconstruction quality.}
    For each example, we show the full frame, a cropped
    region, and the corresponding spatiotemporal slice
    along the marked horizontal line.
    BPP / DISTS / PSNR are reported for each method.
    SGVC achieves competitive perceptual quality while
    preserving favorable spatial fidelity and temporal
    coherence.}
    \label{fig:supp_qualitative}
\end{figure*}

\section{Additional Experimental Results}
\label{sec:supp_experiment}
\paragraph{Evaluation Settings}
Following GLC~\cite{qi2025glc}, we compute FID using
$256\times256$ patches extracted from ground-truth and
reconstructed frames. For the rate--distortion--perception
comparisons in the main paper, all methods are evaluated
with an intra period of 32. For the additional comparisons
in \cref{fig:supp_longseq_rd}, we further evaluate the first 96
frames of each sequence on HEVC Class B and UVG with
a single I-frame (intra period of $-1$), covering a broader
range of fidelity, perceptual, and temporal consistency
metrics.

\paragraph{Quantitative Results}
As shown in \cref{fig:supp_longseq_rd} and discussed in the preceding section, SGVC exhibits mild perceptual degradation over long prediction chains but maintains competitive LPIPS and DISTS performance. Notably, SGVC achieves favorable FID results and higher PSNR than YODA and GVC-RT~\cite{dang2026gvcrt}, consistent with the results in the main paper. GVC-1D~\cite{Zheng_2026_CVPR} incorporates additional long-term context modeling and achieves strong perceptual performance over extended prediction chains, while SGVC remains competitive without a dedicated long-term memory module. While FloLPIPS and warping error $E_{\mathrm{warp}}$ reflect temporal consistency, they are also influenced by spatial quality. We therefore complement these metrics with qualitative comparisons to further assess temporal reconstruction quality.

\paragraph{Qualitative Results}
\Cref{fig:supp_qualitative} presents additional qualitative comparisons using spatial crops and spatiotemporal slices extracted along the marked horizontal lines. Compared with GVC-1D, YODA, and GVC-RT, SGVC better preserves the original structures and introduces fewer implausible details, while maintaining competitive perceptual quality. The spatiotemporal slices further illustrate temporal coherence in the reconstructed sequences, although some inconsistencies remain in challenging long sequences.

\begin{figure}[t]
\centering
\includegraphics[width=\linewidth]{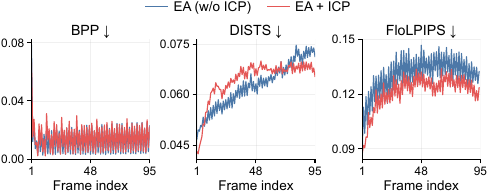}
\caption{\textbf{Ablation of ICP over extended prediction chains on UVG.} Per-frame metrics are averaged across all sequences.}
\label{fig:supp_icp_long}
\end{figure}

\paragraph{Ablation of ICP on Long Sequences}
We further evaluate the effect of In-Context Propagation (ICP) over extended prediction chains on UVG, using 96 frames without periodic intra refresh. Both variants are trained for 14,000 steps, followed by 1,000 steps of fine-tuning. As shown in Fig.~\ref{fig:supp_icp_long}, both variants maintain stable bitrates without systematic growth. With EA enabled, ICP achieves lower FloLPIPS across most frame positions while maintaining comparable DISTS performance at similar bitrates.

\section{Complexity Analysis}
\label{sec:supp_complexity}
The ablation experiments in the main paper use unmerged
LoRA adapters, while all models evaluated in this
section use merged LoRA weights for inference.

\begin{table}[t]
\centering
\footnotesize
\setlength{\tabcolsep}{2pt}
\renewcommand{\arraystretch}{1.12}

\caption{\textbf{Computational complexity at different resolutions.}
Coding times are averaged over P-frames, while memory
denotes peak allocated GPU memory.}
\label{tab:supp_complexity}

\begin{tabular*}{\linewidth}{
@{\extracolsep{\fill}}lrrrrrr@{}}
\toprule
& \multicolumn{2}{c}{kMACs/pixel $\downarrow$}
& \multicolumn{2}{c}{Time (ms) $\downarrow$}
& \multicolumn{2}{c}{Mem. (GiB) $\downarrow$} \\
\cmidrule(lr){2-3}
\cmidrule(lr){4-5}
\cmidrule(l){6-7}
Method & Enc. & Dec. & Enc. & Dec. & Enc. & Dec. \\
\midrule

\multicolumn{7}{@{}l}{
\textit{Class B ($1920\times1080$)}} \\
S$^2$VC & 9700 & 9175 & 1563.7 & 1381.6 & 8.88 & 8.85 \\
YODA & 10490 & 6901 & 1408.7 & 885.0 & 11.41 & 11.38 \\
\textbf{SGVC (ours)}
& \textbf{227} & \textbf{5191}
& \textbf{117.2} & \textbf{845.7}
& \textbf{3.01} & \textbf{5.51} \\
\midrule

\multicolumn{7}{@{}l}{
\textit{Class E ($1280\times720$)}} \\
S$^2$VC & 9194 & 8638 & 676.5 & 595.1 & 5.61 & 5.60 \\
YODA & 11121 & 7320 & 689.8 & 452.7 & 6.41 & 6.40 \\
\textbf{SGVC (ours)}
& \textbf{226} & \textbf{5097}
& \textbf{52.8} & \textbf{352.3}
& \textbf{2.56} & \textbf{3.66} \\
\midrule

\multicolumn{7}{@{}l}{
\textit{Class C ($832\times480$)}} \\
S$^2$VC & 8651 & 8096 & 319.7 & 257.1 & 3.97 & 3.97 \\
YODA & 11154 & 7353 & 352.1 & 250.5 & 3.89 & 3.88 \\
\textbf{SGVC (ours)}
& \textbf{226} & \textbf{5016}
& \textbf{26.7} & \textbf{142.8}
& \textbf{2.36} & \textbf{2.84} \\

\bottomrule
\end{tabular*}
\end{table}

\paragraph{Computational Efficiency}
\Cref{tab:supp_complexity} compares the computational
complexity of SGVC with recent diffusion-based
video compression methods, including S$^2$VC
and YODA. SGVC achieves the lowest
computational cost, coding latency, and peak GPU
memory consumption across all evaluated resolutions.
At $1920\times1080$, SGVC requires
227 kMACs/pixel and 117.2 ms for P-frame encoding,
reducing encoding time by 92.5\% and 91.7\%
compared with S$^2$VC and YODA, respectively.
For decoding, SGVC reduces latency by 38.8\%
and 4.4\%, respectively, while requiring
5.51 GiB of peak GPU memory.

\paragraph{Model Complexity}
\Cref{tab:supp_parameters} reports the parameter counts of SGVC. The I-frame and P-frame models contain 1.18B and 1.15B parameters, respectively, of which only 46.02M and 22.64M belong to their contextual codecs.

\begin{table}[t]
\centering
\footnotesize
\setlength{\tabcolsep}{3pt}
\renewcommand{\arraystretch}{1.12}

\caption{\textbf{Parameter count of SGVC.}
Total parameter counts are reported separately for
the I-frame and P-frame models with merged LoRA weights.
LoRA counts are reported before merging and are not
added to the totals.}
\label{tab:supp_parameters}

\begin{tabular*}{\linewidth}{
@{\extracolsep{\fill}}lrr@{}}
\toprule
Module & I-frame (M) & P-frame (M) \\
\midrule
Codec & 46.02 & 22.64 \\
\quad Feature Adapter & 0.37 & 5.39 \\
\addlinespace[2pt]
DiT Refinement & 1132.34 & 1132.34 \\
\quad Patch Embedder & 1.18 & 1.18 \\
\quad Residual Head & 4.50 & 4.50 \\
\quad LoRA (pre-merge) & 63.90 & 63.90 \\
\midrule
\textbf{Total} & \textbf{1178.37} & \textbf{1154.98} \\
\bottomrule
\end{tabular*}
\end{table}

\section{Limitations}
Although SGVC supports causal frame-by-frame decoding without clip-wise buffering, its decoding complexity remains relatively high due to the large pretrained DiT. At $1920\times1080$, P-frame decoding requires $845.7$ ms per frame ($1.18$ fps) on an NVIDIA RTX 3090, limiting real-time deployment. Moreover, our long-sequence fine-tuning relies on temporally reversed and concatenated short clips rather than genuine long videos. While this strategy mitigates error accumulation, some perceptual degradation persists over extended prediction chains. 
Future work will explore more efficient generative refinement architectures to reduce decoding latency and curate a high-quality dataset of genuine long video sequences to improve temporal robustness.

\end{document}